\documentclass[preprint2,emulateapj5]{aastex}
\usepackage{lscape}
\usepackage{longtable}
\usepackage{amssymb}
\usepackage{apjfonts}

\begin{document}

\title{Dusty Debris Disks as Signposts of Planets: Implications for SIRTF}
\author{B. Zuckerman }
\email{ben@astro.ucla.edu}

\and{}

\author{Inseok Song}
\email{song@astro.ucla.edu}

\affil{Department of Physics and Astronomy\\
Center for Astrobiology\\
University of California, Los Angeles\\
Los Angeles, CA 90095--1562, USA\\
}

\begin{abstract}
Submillimeter and near-infrared images of cool dusty debris disks
and rings suggest the existence of unseen planets. At dusty but
non-imaged stars, semi-major axes of associated planets can be
estimated from the dust temperature. For some young stars these
semi-major axes are greater than an arc second as seen from Earth.
Such stars are excellent targets for sensitive near-infrared
imaging searches for warm planets.  To probe the full extent of
the dust and hence of potential planetary orbits, SIRTF
observations should include measurements with the 160$\,\mu$m
filter. 
\end{abstract}


\section{Introduction}

At near-infrared wavelengths, adaptive optics on 8~m class ground-based
telescopes and the NICMOS camera on the Hubble Space Telescope can
probe regions within a few arc seconds of nearby stars. In such regions,
for stars younger than $\sim 100$~Myrs, warm massive planets can
be detected. Dozens of such young stars within $\sim 60$~pc of Earth
have been identified \citep[and references therein]{SZB03,ARAA}.

Extant images at submillimeter and near-infrared wavelengths of
cool dusty debris disks at main sequence stars (the so-called Vega
phenomenon) usually show substantial spatial structure (e.g.,
\citealt{Holland98} \& \citeyear{Holland03};
\citealt{Greaves,Schneider,Krist,Koerner,Wilner};
\citealt{Weinberger02} \& \citeyear{Weinberger03};
\citealt{Wahhaj,Clampin,Zuckerman01}, and references therein).
Specifically, the dusty regions around $\epsilon $~Eri, Vega,
Fomalhaut, $\beta $~Pic, and HD~141569 all show obvious
non-axisymmetric structure. HR~4796 is orbited by a narrow dusty
ring. Only the dust at TW Hya has, so far, failed to reveal any
structure of note.

Excepting perhaps HD~141569 \citep[and references
therein]{Clampin}, the most likely explanation of the observed
structures are gravitational perturbations by planets with
semi-major axes comparable to the radius of the dusty rings and
disks. For Vega, a 3 Jupiter mass ($M_{J}$) planet is suggested
\citep{Wilner}, while for $\epsilon $~Eri either a 0.2~$M_{J}$
\citep{Ozernoy} or a 0.1~$M_{J}$ \citep{Quillen} planet has been
proposed. Indeed, COBE found that Earth is led and trailed in its
orbit around the Sun by clumps of dust particles \citep{COBE}.
Additional discussions of planet/disk interactions can be found in
\citet{Holland03}, \citeauthor{Kenyon} (2002a, 2003), \citet{Kuchner},
\citet{WD02}, and \citet{Wyatt03}.

Alternative mechanisms to generate structure in the dusty disks of
other stars have been proposed, but likely account for very few,
if any, of the observed structures. \citet{Kalas} investigated the
possibility that a recent, close encounter with a passing star
generated the structure observed in the northeast arm of the
$\beta $~Pic disk. While such an encounter might occur on rare
occasions, such low-probability events cannot plausibly generate
dusty structure in a high percentage of the Vega-like stars
\citep[e.g.,][]{KB02}.
\citet{Takeuchi} proposed that ring structure, such as is observed
at HR~4796 could be generated by dust particles migrating in an
optically thin gaseous disk.  This mechanism is implausible as an
explanation for most, perhaps all, of the observed dusty
structures because of (1)\,their often dramatic non-axisymmetric
shape and (2)\,no evidence of gas at a majority of the above
listed stars including even the model prototype HR~4796
\citep[e.g., ][and references therein]{Zuckerman01}.

As a consequence of the above observations and arguments, one may
reasonably assume that most stars with imaged dust emission have
at least one planet on a wide orbit (tens of AU). From this
assumption it follows that at stars with photometrically detected
(but not yet imaged) dust there is a high probability of orbiting
planets including at least one with a semi-major axis comparable
to that of the dimensions of the dust disk.

Extrapolating from the preceding, one might speculate that young
stars that do not have dust detectable by SIRTF do not possess
planets on wide orbits (see discussion at end of Section~2).
Although dust has not yet been measured directly in the Sun's
Kuiper Belt, dust is very likely present and its presence is
consistent with the existence of a planet (Neptune) at a roughly
comparable distance from the Sun.  Indeed, as has been pointed out
by numerous researchers, with but few exceptions
\citep[e.g.,][]{Chen}, the dusty disks detected around main
sequence stars (by IRAS and ISO) are likely analogs of the Kuiper
Belt.

Twenty years ago, when IRAS first discovered the Vega-phenomenon,
astronomers wondered whether a substantial link connected the
presence of dust with planets. While stars with imaged dust remain
frustratingly few in number, the common and striking structures
that are seen now suggest that the link is substantial and that
dusty stars represent excellent targets for planet searches.

For main sequence stars with circumstellar dust detected by IRAS
and/or ISO, the measured dust temperature is an indication of how
far from a star the dust is located. In the following section we
outline how the dust can be used to guide imaging planet searches
and list specific stars at which to search.

\section{Searching for planets at dusty stars}

The IRAS database has been the prime source for identification of
dusty main sequence stars. A few additional stars were added by
ISO studies. A list of surveys for circumstellar dust appears in
Section~3 of \citet{Zuckerman01} who remarked that, within certain
constraints, the PhD thesis of Murray \citet{Silverstone}
represents the most comprehensive search to date of the IRAS
catalogs for Vega-like stars.

In the discussion that follows and in Table~1, we focus primarily
on stars listed in Silverstone's thesis. One of his constraints
was that IRAS detected excess emission (due to the presence of
dust) at 60~$\mu $m wavelength. Thus, all stars in Table~1 are
orbited by some dust that is sufficiently warm to radiate
significantly at 60~$\mu $m.  Table~1 also includes HIP~13402 and
71284 from \citet{Habing}.  Not all IR excess stars in Silverstone
(2000) appear in our Table~1. For example, we excluded stars
beyond 100\,pc from Earth and stars with small dust optical depth
$\tau$ (see below for definition of $\tau$). We examined each
putative excess star on the Digital Sky Survey plates and
eliminated a few where it appears likely that the apparent far-IR
stellar excess was instead emission from a galaxy near the star in
the plane of the sky. We also have not included the big three
(Vega, Fomalhaut, and $\beta$\,Pic) in Table~1. We felt that these
have been analyzed and imaged to death already and we have nothing
to add. The same might be said about $\epsilon$\,Eri, but we
included it in Table~1 because it well illustrates the
conservative nature of the entries in columns 9 \& 10 of Table~1
(see discussion below).

For Table~1 stars, we have constructed spectral energy
distributions (SEDs) using optical, 2MASS, IRAS, and, when
available, ISO fluxes.  Synthetic stellar spectra of
\citet{Hauschildt}, $Z=0.02$ and $\log g=4.5$ model, are used in
our SED fit process.  Dust emission model parameters were found
from eye-fitting model dust emission to IRAS (and ISO for some
cases) measured fluxes (Figures~1--8).  By assuming that the dust
particles radiate like black bodies at temperature $T_{dust}$, one
may derive their characteristic orbital radii ($R_{dust}$).  These
are indicated in column 9 of Table~1, in Astronomical Units.  Also
indicated in Table~1 are the apparent angular radii (column 10)
that characterize the dust distributions as seen from Earth.

$R_{dust}$ was calculated from 

\[
R_{dust}=\left(\frac{R_{star}}{2}\right)\left(\frac{T_{star}}{T_{dust}}\right)^{2}\]

\noindent where $R_{star}$ and $T_{star}$ are given in columns 6
and 7 of Table~1 and are obtained from the SED fit and the
distance $D$ between Earth and star (column 5). We can check the
accuracy of this technique directly for HIP~8102 ($=\tau $~Ceti)
for which the VLT interferometer measured a stellar radius of
0.773~$R_{\odot }$ \citep{Pijpers}; our technique yields
0.79~$R_{\odot }$. An independent team of investigators
(Kervella~et~al. 2003), who also use the VLT interferometer,
report a radius of 0.804~$R_{\odot }$ for HIP~8102.  In addition,
they measure 0.738~$R_{\odot }$ for HIP~16537 ($\epsilon$~Eri);
our technique yields 0.70~$R_{\odot }$.

For many of the listed stars, these $R_{dust}$ are
``conservative'' in that substantial amounts of dust might exist
at larger distances.  For example, particles at the IRAS color
temperature that are small and radiate less effectively than black
bodies in the far-IR will be located further from the star than
indicated in Table~1. Also, particles that are too cold to radiate
much at 60~$\mu $m would not have been detected by IRAS.

\citet{Z93} deduced that, for the A-type stars Vega, Fomalhaut and
$\beta $~Pic, there is not much dust too cold to have been
detected by IRAS. This conclusion was confirmed by
\citet{Holland98} and by \citet{Harvey}. In addition,
Holland~et~al. noted that the dominant radiating particles at
Fomalhaut appear to be behaving like black bodies.

In contrast, at some stars there is substantial evidence
(summarized in Section 4.2 of Zuckerman 2001) for particles at
larger distances than implied by the IRAS color temperature and
the blackbody assumption.  The characteristic dust orbital radius
(19~AU) calculated (Table~1) from the IRAS color temperature for
the K2 star $\epsilon $~Eri (HIP~16537) is a few times smaller
than the radius of the dust ring imaged by \citet{Greaves} at
800~$\mu $m wavelength. Should dust be present $\gtrsim 100$~AU
from late-type stars, then such stars are insufficiently luminous
to heat all surrounding dust to the $\sim 30$~K required for
generation of substantial 60~$\mu $m emission. For example, a G0
star can heat large (black-body) particles to $\sim 30$ K at a
distance of 100~AU. The corresponding distances for K0 and M0
stars are only $\sim 65$ and 25~AU, respectively.

In addition, even for some early-type stars, our expression for
R$_{dust}$ will underestimate the true extent of the dust
distribution.  For example, at HR~4796, our SED fit implies
R$_{star}$, T$_{star}$, and T$_{dust}$ of 1.86\,R$_{\odot }$,
9200\,K, and 100\,K respectively, which yields a calculated
R$_{dust}=37$\,AU.  However, the radius of the dusty ring imaged by
NICMOS with the Space Telescope is $\sim65$\,AU (see, e.g.,
Figure~1 in Zuckerman 2001).

In column 11 of Table~1,  $\tau $ is the total energy emitted by
dust grains divided by the bolometric luminosity of the star.
$\tau $ is a measure of the fraction of the ultraviolet and visual
light emitted by the star that is absorbed by the orbiting dust
particles. The listed values of $\tau $, which are obtained from
the SEDs shown in Figures~1--8, usually agree reasonably well with
values of $\tau $ given in Silverstone (2000), although
discrepancies of a factor of two or three sometimes appear.  $\tau
$ may be used as one age indicator. From \citet{Spangler}, if
$\tau \gtrsim $0.001, then a star is probably not older than $\sim
100$~Myrs.  It may be seen that most stars in Table~1 with
multiple age indicators conform to this rule, but there are a few
potential exceptions, notably HIP~35457, 69682, \& 87815.

Additional techniques for estimating stellar ages are listed in
\citet{bPic2} and \citet{SZB03}. We used a variety of techniques,
indicated in column 13 of Table~1, to deduce the ages in column
12. Some of the ages are quite secure, others not so secure (one
question mark), or quite uncertain (two question marks).

Planet detection with current adaptive optics (AO) imaging systems
on large telescopes requires planet-star separations of at least
one to two arc seconds. In addition, a planet must be sufficiently
warm to radiate at near-IR wavelengths. Long integrations on
nearby stars with ages of hundreds of millions of years can probe
down to a few Jupiter masses (e.g., \citealt{Macintosh}). Ages of
tens of millions of years or less are required if planets of a
Jupiter mass are to be detected.  Thus, to optimize imaging planet
searches accurate stellar ages are required. 

Because AO (and HST/NICMOS) sensitivities are telescope,
elevation, and wavelength specific and because H, K$'$,
and L$'$ thermal fluxes from planets with temperatures
$\lesssim700$\,K are based on model predictions and not on
observations, it is not possible to give a general prescription
regarding which stars in Table~1 should be observed with AO
and at what wavelength. Indeed an AO system has been
commissioned on the VLT only within the past year and Gemini
North and South and Subaru have had no or only rudimentary AO
systems.                                                                     
                                                                                
That said, for a given star/dust angular separation (column 10 in
Table~1), planet detectability is maximized by observing the
youngest, closest stars to Earth.  For example, a two
Jupiter mass planet will fade by 5-6 magnitudes, in an
absolute sense and also relative to the brightness of its
star, at H and K bands as it ages from 10 to 100\,Myrs. Similarly,
a planet 50\,pc from Earth will be 3.5\,magnitudes fainter
than a comparably warm one only 10\,pc away (while the
star/planet contrast is independent of distance).  Then there
is stellar spectral type to consider.  A planet of given
mass, age, and semi-major axis will be harder to detect close to
an intrinsically luminous A-type star compared to one of
K-type.  But if more massive planets, perhaps with larger
semi-major axes, form near the relatively more massive
stars, then this might more than compensate for         the
unfavorable contrast ratio.  In any event, because AO systems
(unlike      HST/NICMOS) are not well suited to detection of
extended objects, near          infrared light scattered by dust,
even for stars with large $\tau$, is very        unlikely to
hamper planet detection.

Both observation \citep{Spangler} and theory suggest that, on
average, young stars will have a dustier ``Kuiper Belt'' than old
stars. Based on the mass of Neptune, the mass ($\sim3\,M_{Neptune}$) of
the proposed planet on a wide orbit at $\epsilon$\,Eri, and the
decrease of $\tau$ with time indicated in Figure~2 of
\citet{Spangler}, a star of age $\lesssim$ few 100 million years
with a similarly massive planet on a wide orbit would have
$\tau\gtrsim10^{-5}$. SIRTF should generally be capable of
detecting $\tau$ as small as $10^{-6}$. Thus, if the dust-planet
connection is as strong as we suggest, then Jupiter mass planets
on wide orbits will rarely, if ever, be detected arround young
stars that lack SIRTF detected far-IR excess emission. The same
may be true for Saturn or even Neptune mass planets, although
these, of course, will be much harder to detect.

\section{Conclusions}

Recent imaging observations and analysis of structure in the dusty
debris disks that surround a handful of nearby stars suggest that
they possess planets on wide orbits (tens of AU). Thus, dusty stars,
whether imaged or not, are excellent targets for planet searches.
Young stars with cool dust make the best targets for adaptive optics
and HST imaging programs. Older stars and those with somewhat warmer
dust might be better investigated with precision radial velocities.
While the earliest searches of the IRAS catalogs revealed mostly A-type
Vega-like stars which are not suitable for study by the precision
radial velocity technique, a majority of stars in Table 1 are of later
spectral types.

Looking toward the future, one anticipates that SIRTF will search
for dusty debris at all nearby stars that have been identified as
very young (e.g., Zuckerman et al 2001a, Song et al 2003). Since the
presence of cool dust would point toward the existence of planets
on wide orbits accessible to adaptive optics, it is important that
the SIRTF programs include measurements with the 160~$\mu $m filter.

\acknowledgements{
We thank Dr. Murray Silverstone for information regarding the
relative positions of IRAS 12 \& 60\,$\mu$m sources, and Drs. M.
Jura, E. Becklin, and B. Macintosh for helpful comments.
This research was supported by a NASA grant to UCLA and by NASA's
Astrobiology Institute. }



\onecolumn
\begin{figure}
{\centering \resizebox*{0.95\columnwidth}{!}{\includegraphics{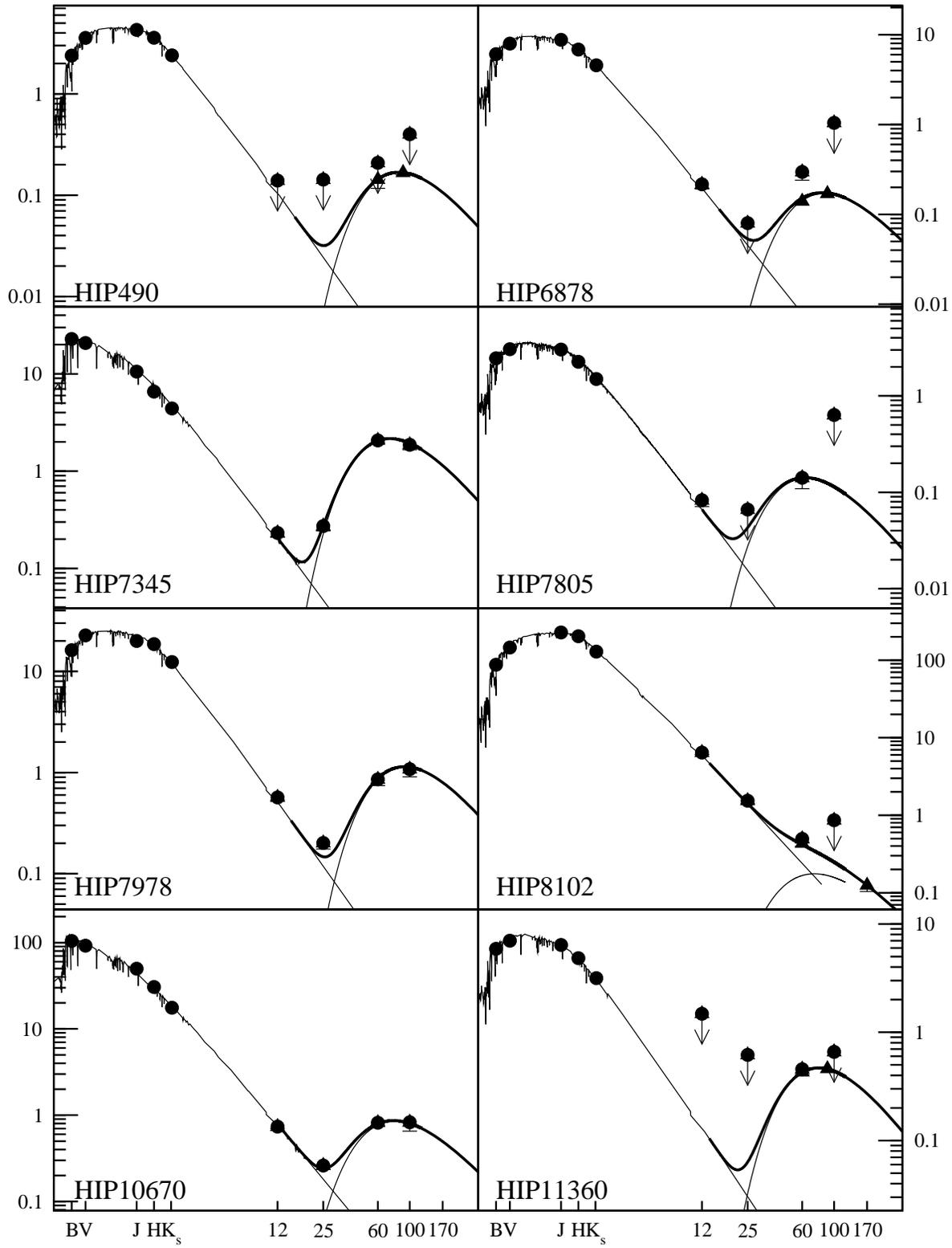}} \par}
\caption{Spectral energy distributions of stars with dusty debris
disks. B \& V fluxes from Hipparcos, $JHK_s$ from 2MASS,
12/25/60/100\,$\mu$m from IRAS. In addition, 60/90/100/170\,$\mu$m
fluxes (triangles) from ISO.}
\end{figure}
\begin{figure}
{\centering \resizebox*{0.95\columnwidth}{!}{\includegraphics{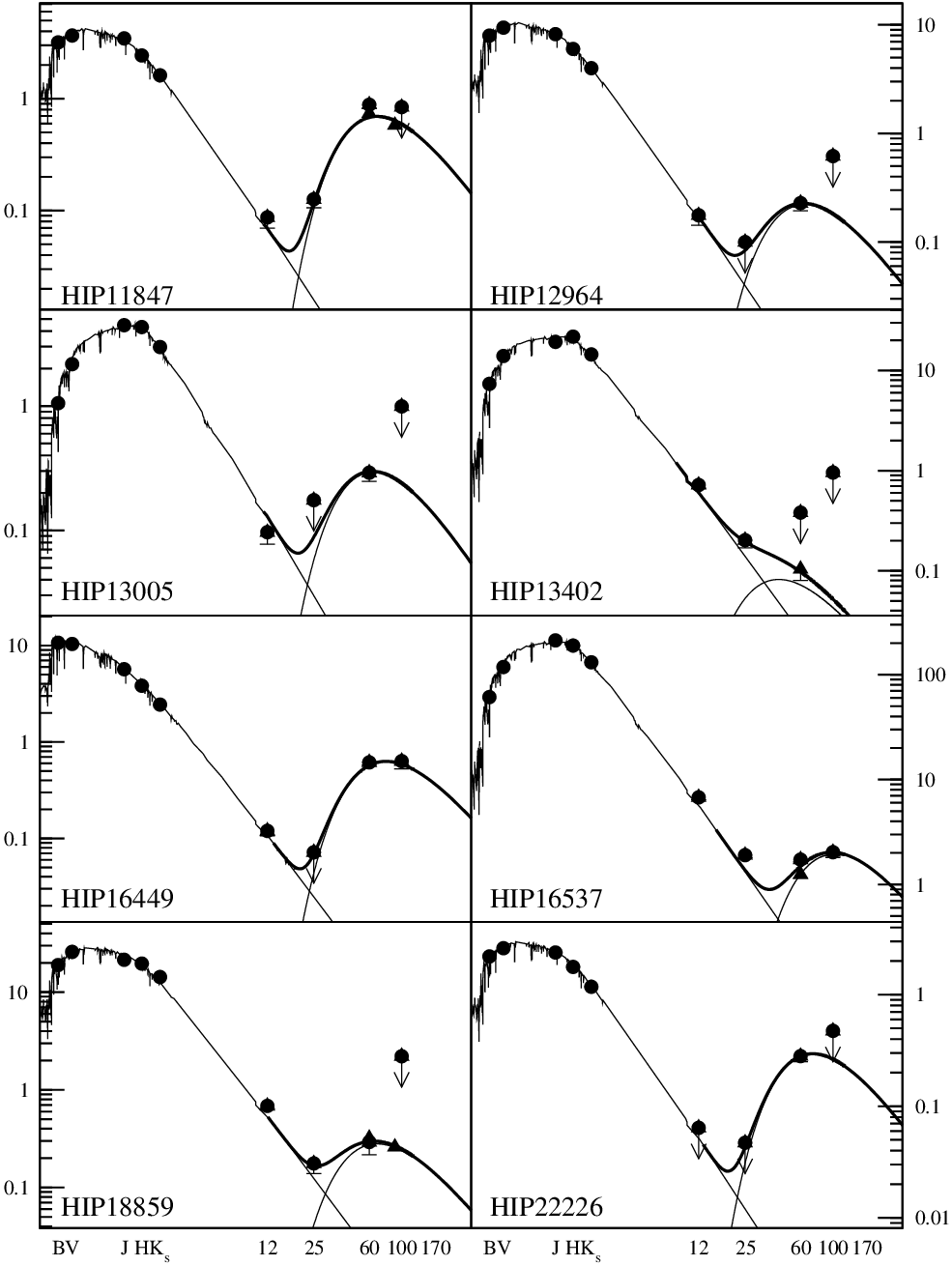}} \par}
\caption{Spectral energy distributions of stars with dusty debris disks
         (See caption to Figure~1)}
\end{figure}
\begin{figure}
{\centering \resizebox*{0.95\columnwidth}{!}{\includegraphics{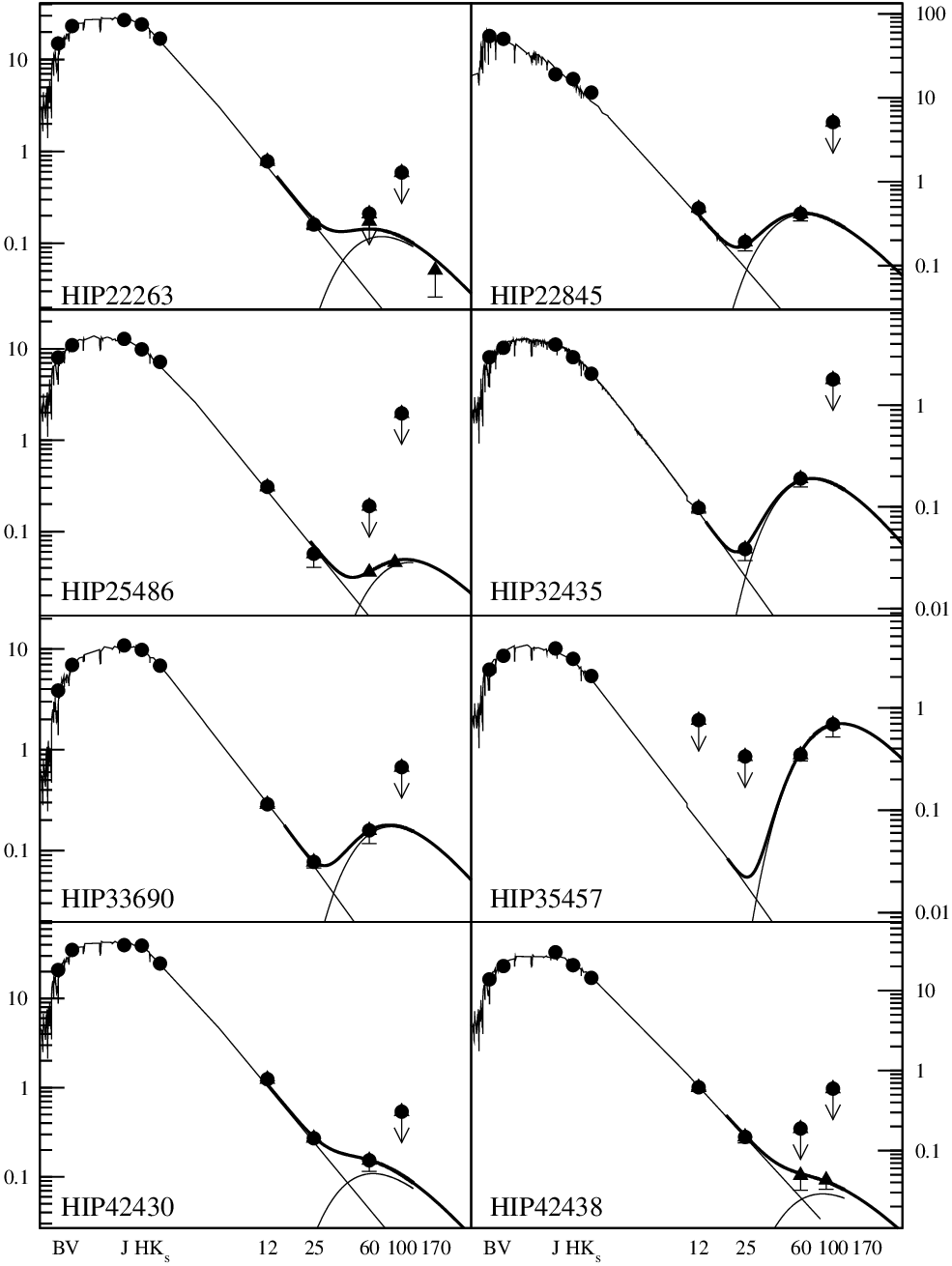}} \par}
\caption{Spectral energy distributions of stars with dusty debris disks
         (See caption to Figure~1).}
\end{figure}
\begin{figure}
{\centering \resizebox*{0.95\columnwidth}{!}{\includegraphics{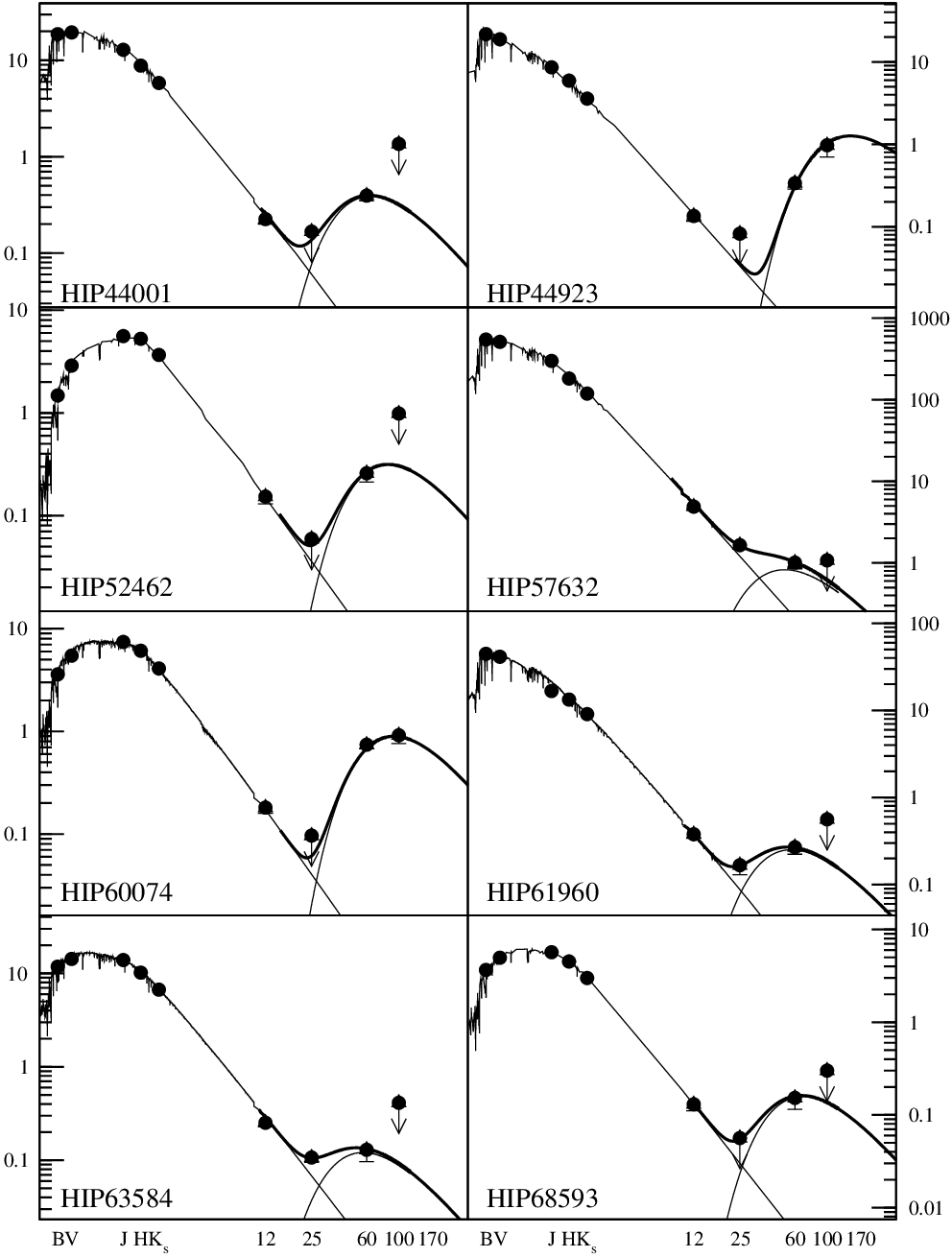}} \par}
\caption{Spectral energy distributions of stars with dusty debris disks
         (See caption to Figure~1).}
\end{figure}
\begin{figure}
{\centering \resizebox*{0.95\columnwidth}{!}{\includegraphics{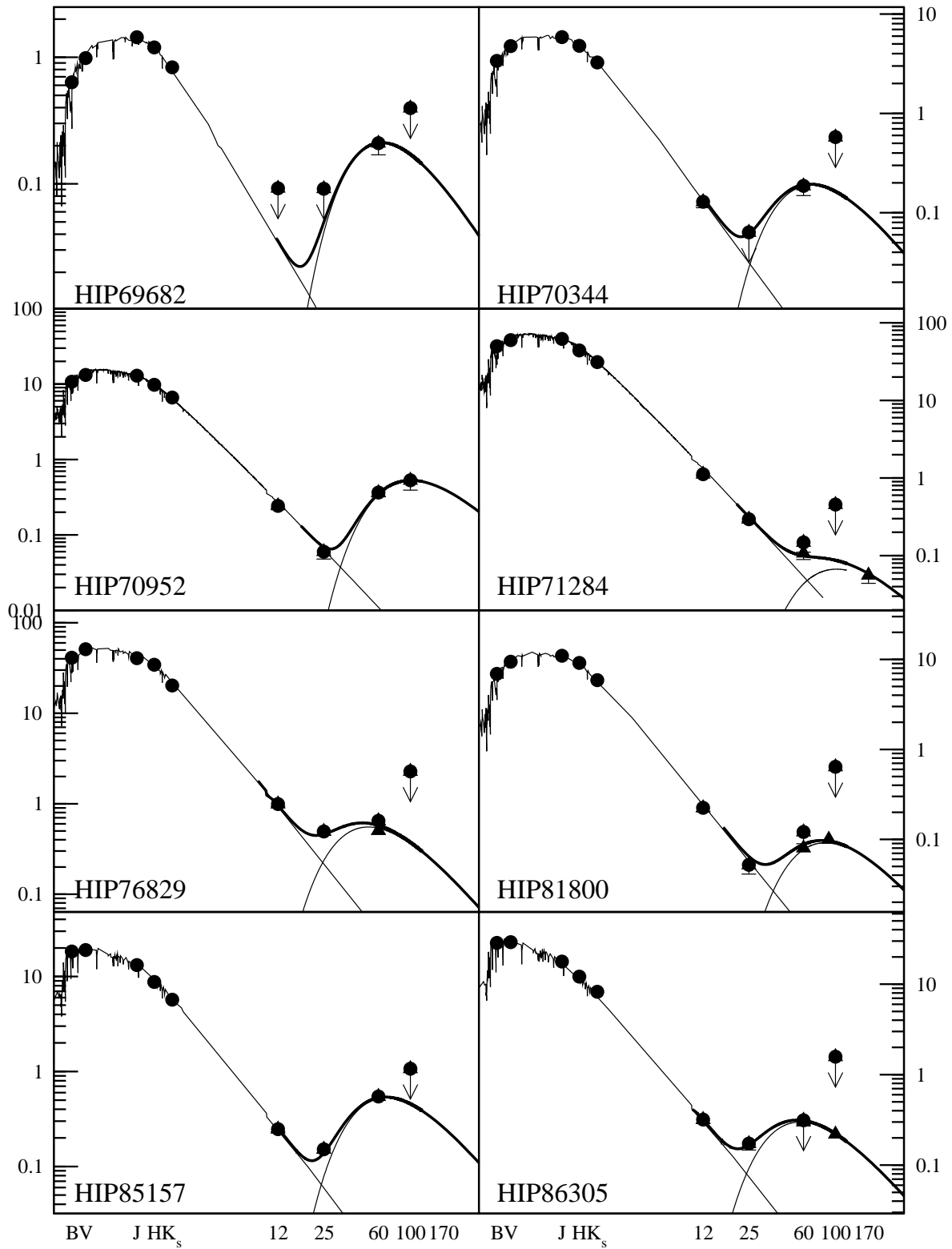}} \par}
\caption{Spectral energy distributions of stars with dusty debris disks
         (See caption to Figure~1). 60 and 100\,$\mu$m fluxes (triangles) 
         for HIP~86305 from \citet{Silverstone}.}
\end{figure}
\begin{figure}
{\centering \resizebox*{0.95\columnwidth}{!}{\includegraphics{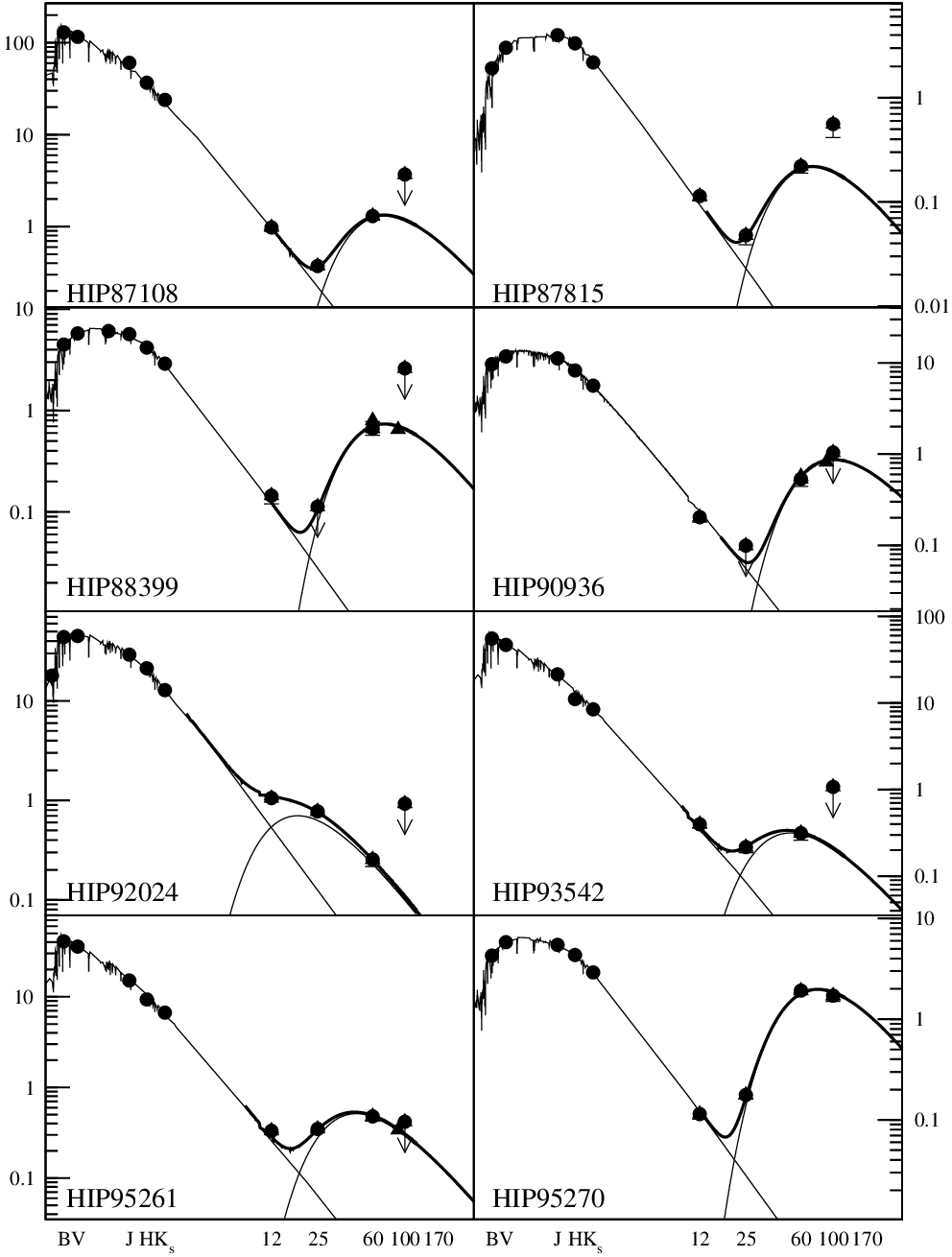}} \par}
\caption{Spectral energy distributions of stars with dusty debris disks
         (See caption to Figure~1).}
\end{figure}
\begin{figure}
{\centering \resizebox*{0.95\columnwidth}{!}{\includegraphics{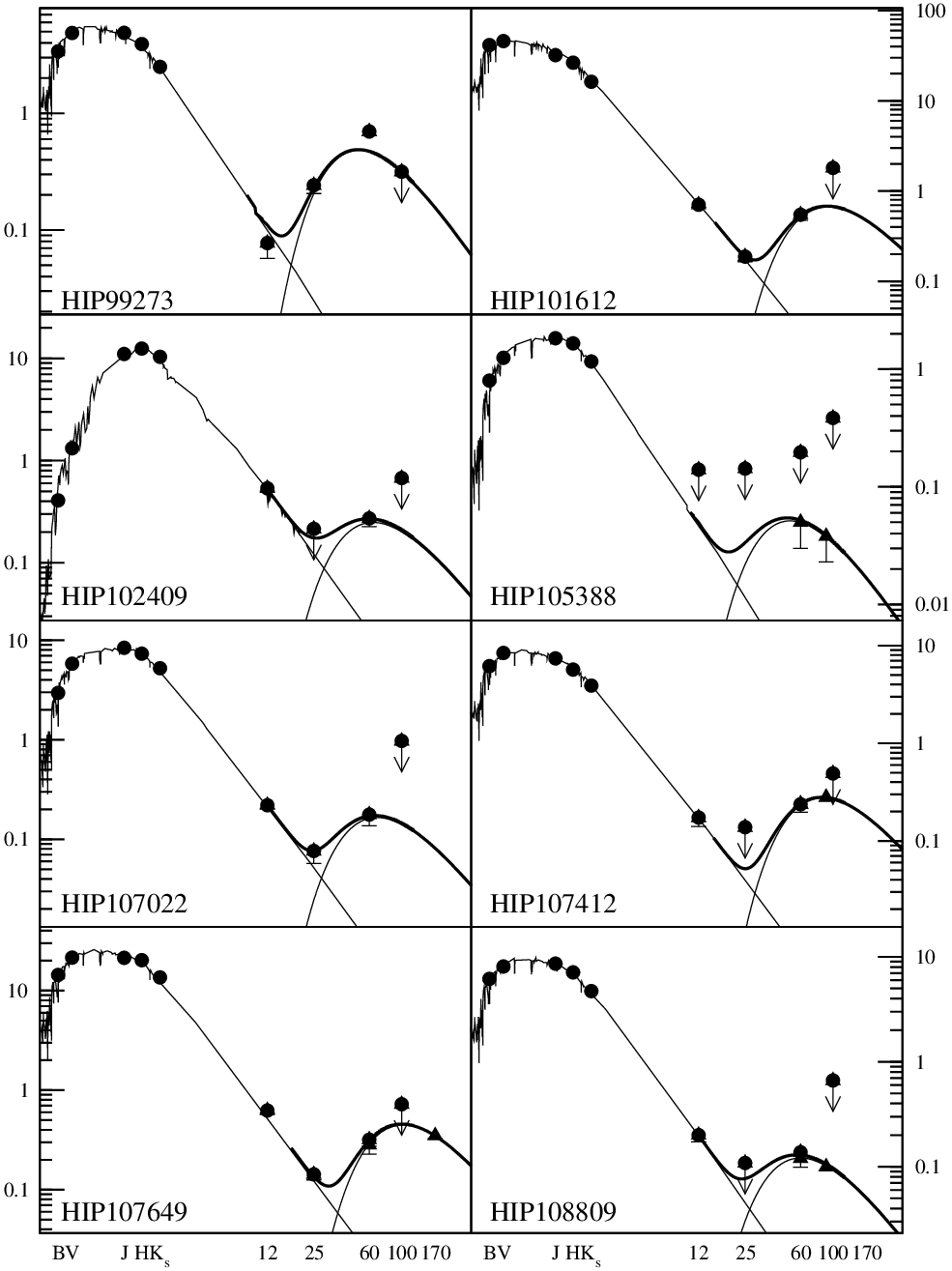}} \par}
\caption{Spectral energy distributions of stars with dusty debris disks
         (See caption to Figure~1).}
\end{figure}
\begin{figure}
{\centering \resizebox*{0.95\columnwidth}{!}{\includegraphics{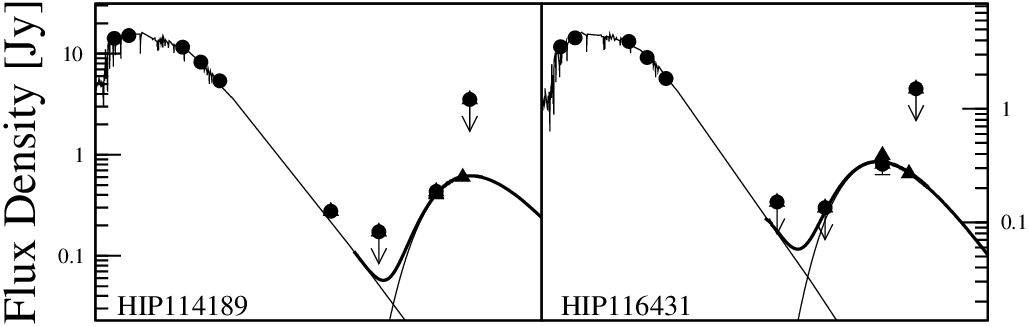}} \par}
\caption{Spectral energy distributions of stars with dusty debris disks
         (See caption to Figure~1).}
\end{figure}
\onecolumn
\def\mc{\multicolumn}

\begin{landscape}
 \begin{longtable}{cccccccccccccc}
 \caption{Stars with Dusty Debris Disks} \\
 \hline
 HIP    &  HD    & Sp.  &   V   &   D  &$R_{star}$&$T_{star}$&$T_{dust}$&$R_{dust}$& angle & $\tau$ & age & age method & notes \\
        &        & Type & [mag] & [pc] &[$R_\odot$]& [K]     &  [K]     & [AU]     & [$''$]&$\times10^{-4}$& [Myrs] &  & \\
  (1)   &  (2)   & (3)  &  (4)  &  (5) &   (6)     & (7)     &  (8)     &  (9)     &   (10)&    (11)       &  (12)  & (13)  & (14) \\
 \hline
 \endfirsthead
 \caption[]{(continued)}\\
 \hline
 HIP    &  HD    & Sp.  &   V   &   D  &$R_{star}$&$T_{star}$&$T_{dust}$&$R_{dust}$& angle & $\tau$ & age & age method & notes \\
 \hline
 \endhead
 \hline
 \multicolumn{14}{r}{continued on next page}\\
 \hline
 \endfoot
 \endlastfoot
 490    & 105    & G0V    & 7.5 & 40.2 & 1.16 & 5800 &  60 &  27 & 0.68  & 5    &$\lesssim100$& a,b,c  &  1        \\
 6878   & 8907   & F8     & 6.7 & 34.2 & 1.34 & 5800 &  60 &  31 & 0.92  & 3    &   200?      & a,b,c  &  1        \\
 7345   & 9672   & A1V    & 5.6 & 61.3 & 1.96 & 8600 &  70 &  74 & 1.21  & 6    &    20?      & ZFK    &  2        \\
 7805   & 10472  & F2IV/V & 7.7 & 66.6 & 1.40 & 6400 &  80 &  22 & 0.34  & 6    &    30       & ZSW    &           \\
 7978   & 10647  & F8V    & 5.5 & 17.4 & 0.99 & 6200 &  55 &  31 & 1.81  & 3    &   300?      & a,b,c  &           \\
 8102   & 10700  & G8V    & 3.6 &  3.6 & 0.79 & 5400 &  70 &  12 & 3.26  & 0.1  &   7000??    & a,b    &  1        \\
 10670  & 14055  & A1Vnn  & 4.0 & 36.1 & 2.09 & 9400 &  65 & 109 & 3.03  & 0.6  &   100?      & a,d    &           \\
 11360  & 15115  & F2     & 6.8 & 44.8 & 1.30 & 6800 &  65 &  36 & 0.79  & 5    &   100?      & a,b,c  &  1        \\
 11847  & 15745  & F0     & 7.5 & 63.7 & 1.30 & 6800 &  75 &  27 & 0.42  & 12   &    30?      & d,e    &  1        \\
 12964  & 17390  & F3IV/V & 6.6 & 45.1 & 1.47 & 6800 &  80 &  27 & 0.59  & 2    &   300??     & a      &           \\
 13005  &  ---   & K0     & 8.1 & 67.7 & 2.22 & 5000 &  60 &  39 & 0.57  & 14   &     ?       & b,e    &  3        \\
 13402  & 17925  & K1V    & 6.0 & 10.4 & 0.73 & 5200 &  60 &  14 & 1.32  & 0.8  &$\lesssim100$& a,b,c  &  1        \\
 16449  & 21997  & A3IV/V & 6.7 & 73.8 & 1.64 & 8400 &  65 &  68 & 0.93  & 3    &   100?      & d      &           \\
 16537  & 22049  & K2V    & 3.7 &  3.2 & 0.70 & 5200 &  50 &  19 & 5.90  & 0.7  &   730       & S2000  &  1        \\
 18859  & 25457  & F5V    & 5.4 & 19.2 & 1.15 & 6400 &  75 &  21 & 1.09  & 0.8  &    30       & a,b,c  &  1        \\
 22226  & 30447  & F3V    & 7.9 & 78.1 & 1.31 & 6600 &  70 &  29 & 0.37  & 11   &$\lesssim100$& e      &           \\
 22263  & 30495  & G3V    & 5.5 & 13.3 & 0.99 & 5600 &  70 &  16 & 1.19  & 0.6  &   300?      & a,b    &  1        \\
 22845  & 31295  & A0V    & 4.6 & 37.0 & 1.56 & 9400 &  80 &  54 & 1.45  & 0.3  &   100?      & a,d    &           \\
 25486  & 35850  & F7V:   & 6.3 & 26.8 & 1.24 & 6000 &  45 &  55 & 2.06  & 0.2  &    12       & ZSBW   &  1        \\
 32435  & 53842  & F5V    & 7.5 & 57.3 & 1.46 & 6200 &  70 &  29 & 0.50  & 4.5  &    30?      & a,b,c  &           \\
 33690  & 53143  & K0IV/V & 6.8 & 18.4 & 0.92 & 5200 &  60 &  17 & 0.94  & 2.5  &   300?      & a,b,c  &           \\
 35457  & 56099  & F8     & 7.6 & 86.8 & 2.17 & 6200 &  45 &  70 & 0.80  & 12   & $>500$?     & a,b,e  &  4        \\
 42430  & 73752  & G3/5V  & 5.0 & 19.9 & 1.83 & 5600 &  80 &  17 & 0.83  & 0.3  & $>600$      & S2000  &  5        \\
 42438  & 72905  & G1.5Vb & 5.6 & 14.3 & 0.99 & 5800 &  60 &  23 & 1.62  & 0.1  &   200?      & a,b,c  &  1        \\
 44001  & 76582  & F0IV   & 5.7 & 49.3 & 1.89 & 7600 &  80 &  43 & 0.85  & 3    &   300??     & a,d    &           \\
 44923  & 78702  & A0/1V  & 5.9 & 79.9 & 2.13 & 9600 &  35 & 400 & 5.00  & 2.5  &   100?      & d      &           \\
 52462  & 92945  & K1V    & 8.0 & 21.6 & 0.81 & 5000 &  60 &  14 & 0.65  & 7    &   100       & SBZ    &           \\
 57632  & 102647 & A3Vvar & 2.1 & 11.1 & 1.81 & 8400 & 100 &  32 & 2.90  & 0.2  &    50       & S2001  &           \\
 60074  & 107146 & G2V    & 7.0 & 28.5 & 1.02 & 5800 &  55 &  28 & 1.00  & 12   &$\lesssim100$&a,b,c,e &           \\
 61960  & 110411 & A0V    & 4.9 & 36.9 & 1.55 & 8800 &  90 &  37 & 1.00  & 0.3  &   100??     & a,d    &           \\
 63584  & 113337 & F6V    & 6.0 & 37.4 & 1.67 & 6600 &  90 &  22 & 0.60  & 1    &    50?      & a,b    &  6        \\
 68593  & 122652 & F8     & 7.2 & 37.2 & 1.18 & 6000 &  75 &  19 & 0.51  & 3    &   300?      & a,b,c  &           \\
 69682  & 124718 & G5V    & 8.9 & 61.3 & 1.03 & 5600 &  80 &  13 & 0.21  & 27   & $>500$      &a,b,c,e &  7        \\
 70344  & 126265 & G2III  & 7.2 & 70.1 & 2.31 & 5800 &  75 &  35 & 0.50  & 2    & $>500$      & a,b    &           \\
 70952  & 127821 & F4IV   & 6.1 & 31.7 & 1.35 & 6600 &  50 &  59 & 1.84  & 1.5  &   200?      & a,b    &           \\
 71284  & 128167 & F3Vvar & 4.5 & 15.5 & 1.46 & 6400 &  50 &  60 & 3.86  & 0.1  &  1000??     & a,b,c  &  1        \\
 76829  & 139664 & F5IV/V & 4.6 & 17.5 & 1.38 & 6600 & 100 &  15 & 0.86  & 0.9  &   200?      & a,b,c  &  1        \\
 81800  & 151044 & F8V    & 6.5 & 29.4 & 1.26 & 6000 &  60 &  32 & 1.09  & 0.8  & $>500$      & a,b    &  1        \\
 85157  & 157728 & F0IV   & 5.7 & 42.8 & 1.63 & 7600 &  75 &  42 & 0.97  & 3    &   100?      & a,d    &           \\
 86305  & 159492 & A7V    & 5.2 & 42.2 & 1.75 & 8200 &  90 &  36 & 0.87  & 1    &   200?      & a,d    &  1        \\
 87108  & 161868 & A0V    & 3.8 & 29.1 & 1.92 & 9400 &  70 &  87 & 2.97  & 0.6  &   200?      & a,d    &           \\
 87815  & 164330 & K0     & 7.7 & 83.2 & 2.30 & 5600 &  70 &  37 & 0.44  & 9    & $>500$      & a,b,e  &           \\
 88399  & 164249 & F5V    & 7.0 & 46.9 & 1.36 & 6400 &  70 &  28 & 0.60  & 19   &    12       & ZSBW   &  1        \\
 90936  & 170773 & F5V    & 6.3 & 36.1 & 1.45 & 6600 &  50 &  63 & 1.75  & 5    &   200?      & a,b,c  &  1        \\
 92024  & 172555 & A7     & 4.8 & 29.2 & 1.61 & 7800 & 280 &   3 & 0.11  & 5    &    12       & ZSBW   &           \\
 93542  & 176638 & A0Vn   & 5.1 & 56.3 & 2.33 & 9800 & 100 &  56 & 1.00  & 0.7  &   200?      & a,d    &           \\
 95261  & 181296 & A0Vn   & 5.0 & 47.7 & 1.63 & 9800 & 110 &  32 & 0.68  & 0.8  &    12       & ZSBW   &  1        \\
 95270  & 181327 & F5.5V  & 7.0 & 50.6 & 1.44 & 6400 &  65 &  35 & 0.68  & 32   &    12       & ZSBW   &           \\
 99273  & 191089 & F5V    & 7.5 & 53.5 & 1.40 & 6400 & 100 &  14 & 0.27  & 13   &$\lesssim100$& a,b,c,e&           \\
 101612 & 195627 & F1III  & 4.5 & 27.6 & 1.86 & 7000 &  55 &  75 & 2.69  & 1    &   200?      & a,d    &           \\
 102409 & 197481 & M1Ve   & 8.8 &  9.9 & 0.76 & 3800 &  80 &   4 & 0.43  & 1    &    12       & ZSBW   &           \\
 105388 & 202917 & G5V    & 8.7 & 45.9 & 0.94 & 5400 & 100 &   7 & 0.15  & 6    &    30       & ZW     &  1        \\
 107022 & 205536 & G8V    & 7.1 & 22.1 & 0.93 & 5800 &  75 &  14 & 0.63  & 3    & $>500$      & a,b    &           \\
 107412 & 206893 & F5V    & 6.9 & 38.9 & 1.30 & 6400 &  60 &  37 & 0.95  & 2    &   200?      & a,b    &  1        \\
 107649 & 207129 & G2V    & 5.6 & 15.6 & 0.98 & 6000 &  50 &  35 & 2.26  & 1.4  &   600       & SZB    &  1        \\
 108809 & 209253 & F6/7V  & 6.9 & 30.1 & 1.09 & 6200 &  85 &  14 & 0.48  & 0.9  &   200??     & a,b,c  &  1        \\
 114189 & 218396 & A5V    & 6.0 & 39.9 & 1.42 & 7400 &  50 &  78 & 1.95  & 1.4  &    30       & a,d    &  1        \\
 116431 & 221853 & F0     & 7.3 & 71.2 & 1.66 & 6600 &  90 &  22 & 0.31  & 12   &$\lesssim100$& e      &  1        \\
 \hline
 \end{longtable}
\end{landscape} 

\noindent Footnotes to Table~1.
\begin{list}{--}{\setlength{\itemsep}{0.2em}}
 \item{Age Methods::}
 \begin{list}{-}{\setlength{\itemsep}{0.2em}}
  \item{S2000=\citet{LateVegas}}
  \item{S2001=\citet{AVegas}}
  \item{SBZ= \citet{MF_TWA}}
  \item{ZFK  =\citet{Z95}      }
  \item{ZSBW=\citet{bPic2} }
  \item{ZSW  =\citet{ZSW}      }
  \item{ZW=  \citet{Z00}   }
  \item{SZB  =\citet{SZB03}    }
  \item{a\hspace{1em}UVW \citep{ARAA}}
  \item{b\hspace{1em}X-ray emission \citep[e.g., ][]{SZB03}}
  \item{c\hspace{1em}lithium age \citep{SZB03}}
  \item{d\hspace{1em}location on an A-star Hertzsprung-Russell diagram \citep{Lowrance}}
  \item{e\hspace{1em}dust (if $\tau\gtrsim10^{-3}$, then age$\lesssim10^8$\,years, \citealt{Spangler})}
 \end{list}
 \item{Calculations use 1\,AU=200\,$R_\odot$}
\end{list}

\newpage
\noindent Notes to Table~1.
\begin{enumerate}
  \item {ISO 60, 90, 100, and/or 170\,$\mu$m fluxes available from \citet{Silverstone} and/or
         \citet{Habing}; see Figures~1--8.}
  \item {HIP~7345 (=49~Cet) is the only known main sequence A-type star with CO emission detected
        with a radio telescope (ZFK), thus suggesting a very young age. But its 
        galactic space motion UVW ($-23,-17,-4$) with respect to
        Sun is not indicative of extreme youth (U is positive
        toward the Galactic center).}
  \item {Based on the offset between the 12 and 60\,$\mu$m IRAS positions, the apparent 
         60\,$\mu$m excess is probably from a galaxy at position angle $\sim70^\circ$ and
         $\sim45''$ from HIP~13005.}
 \item {HIP~35457 is a $0\farcs16$ binary. The very large $\tau$ seems inconsistent with the old
        age implied by the absence of ROSAT All Sky X-ray emission and the stars' galactic
        space motion (UVW).}
 \item {HIP~42430 is a $1\farcs0$ binary.}
 \item {M-star companion LDS\,2662 is very young based on its location on an $M_K$ versus 
        $V-K$ color magnitude diagram (e.g., Figure~2 in SZB).}
 \item {The Galactic space motion (UVW) and absence of lithium and of X-ray emission all
        point to an old star. There is no evidence on the Digital Sky Survey and 2MASS All Sky
        QuickLook Images ($JHK_s$) of a nearby galaxy. Yet $\tau$ is very large.}
\end{enumerate}


\begin{thebibliography}{37}
\expandafter\ifx\csname natexlab\endcsname\relax\def\natexlab#1{#1}\fi

\bibitem[\protect\astroncite{{Chen} \& {Jura}}{2001}]{Chen}
{Chen}, C.~H. \& {Jura}, M. 2001, {\em \apjl\/}, {\bf 560}, L171

\bibitem[\protect\astroncite{{Clampin} {\em et~al.\/}}{2003}]{Clampin}
{Clampin}, M., {\em et~al.\/} 2003, {\em \aj\/}, {\bf 126}, 385

\bibitem[\protect\astroncite{{Greaves} {\em et~al.\/}}{1998}]{Greaves}
{Greaves}, J.~S., {\em et~al.\/} 1998, {\em \apjl\/}, {\bf 506}, L133

\bibitem[\protect\astroncite{{Habing} {\em et~al.\/}}{2001}]{Habing}
{Habing}, H.~J., {\em et~al.\/} 2001, {\em \aap\/}, {\bf 365}, 545

\bibitem[\protect\astroncite{{Harvey} \& {Jefferys}}{2000}]{Harvey}
{Harvey}, P.~M. \& {Jefferys}, W.~H. 2000, {\em \apj\/}, {\bf 538}, 783

\bibitem[\protect\astroncite{{Hauschildt} {\em et~al.\/}}{1999}]{Hauschildt}
{Hauschildt}, P.~H., {Allard}, F., {Ferguson}, J., {Baron}, E., \& {Alexander},
  D.~R. 1999, {\em \apj\/}, {\bf 525}, 871

\bibitem[\protect\astroncite{{Holland} {\em et~al.\/}}{1998}]{Holland98}
{Holland}, W.~S., {\em et~al.\/} 1998, {\em \nat\/}, {\bf 392}, 788

\bibitem[\protect\astroncite{{Holland} {\em et~al.\/}}{2003}]{Holland03}
--- 2003, {\em \apj\/}, {\bf 582}, 1141

\bibitem[\protect\astroncite{{Kalas} {\em et~al.\/}}{2001}]{Kalas}
{Kalas}, P., {Deltorn}, J., \& {Larwood}, J. 2001, {\em \apj\/}, {\bf 553}, 410

\bibitem[\protect\astroncite{{Kenyon} \& {Bromley}}{2002a}]{Kenyon}
{Kenyon}, S.~J. \& {Bromley}, B.~C. 2002a, {\em \apjl\/}, {\bf 577}, L35

\bibitem[\protect\astroncite{{Kenyon} \& {Bromley}}{2002b}]{KB02} Kenyon, S.~J.~\&
Bromley, B.~C.\ 2002b, \aj, 123, 1757

\bibitem[Kenyon \& Bromley(2003)]{KB03} Kenyon, S.~J.~\&
Bromley, B.~C.\ 2003, ArXiv Astrophysics e-prints, 9540

\bibitem[Kervella et al.(2003)]{Kervella} Kervella, P.~et al.\
2003, ArXiv Astrophysics e-prints, 9784

\bibitem[\protect\astroncite{{Koerner} {\em et~al.\/}}{2001}]{Koerner}
{Koerner}, D.~W., {Sargent}, A.~I., \& {Ostroff}, N.~A. 2001, {\em \apjl\/},
  {\bf 560}, L181

\bibitem[\protect\astroncite{{Krist} {\em et~al.\/}}{2000}]{Krist}
{Krist}, J.~E., {Stapelfeldt}, K.~R., {M{\' e}nard}, F., {Padgett}, D.~L., \&
  {Burrows}, C.~J. 2000, {\em \apj\/}, {\bf 538}, 793

\bibitem[Kuchner \& Holman(2003)]{Kuchner} Kuchner, M.~J.~\&
Holman, M.~J.\ 2003, \apj, 588, 1110

\bibitem[\protect\astroncite{{Lowrance} {\em et~al.\/}}{2000}]{Lowrance}
{Lowrance}, P.~J., {\em et~al.\/} 2000, {\em \apj\/}, {\bf 541}, 390

\bibitem[\protect\astroncite{{Macintosh} {\em et~al.\/}}{2003}]{Macintosh}
{Macintosh}, B., {Becklin}, E.~E., {Kaisler}, D., {Konopacky}, Q., \&
  {Zuckerman}, B. 2003, {\em \apj\/}, 594, 538

\bibitem[\protect\astroncite{{Ozernoy} {\em et~al.\/}}{2000}]{Ozernoy}
{Ozernoy}, L.~M., {Gorkavyi}, N.~N., {Mather}, J.~C., \& {Taidakova}, T.~A.
  2000, {\em \apjl\/}, {\bf 537}, L147

\bibitem[\protect\astroncite{{Pijpers} {\em et~al.\/}}{2003}]{Pijpers}
{Pijpers}, F.~P., {Teixeira}, T.~C., {Garcia}, P.~J., {Cunha}, M.~S.,
  {Monteiro}, M.~J.~P.~F.~G., \& {Christensen-Dalsgaard}, J. 2003, {\em
  \aap\/}, {\bf 406}, L15

\bibitem[\protect\astroncite{{Quillen} \& {Thorndike}}{2002}]{Quillen}
{Quillen}, A.~C. \& {Thorndike}, S. 2002, {\em \apjl\/}, {\bf 578}, L149

\bibitem[\protect\astroncite{{Reach} {\em et~al.\/}}{1995}]{COBE}
{Reach}, W.~T., {\em et~al.\/} 1995, {\em \nat\/}, {\bf 374}, 521

\bibitem[\protect\astroncite{{Schneider} {\em et~al.\/}}{1999}]{Schneider}
{Schneider}, G., {\em et~al.\/} 1999, {\em \apjl\/}, {\bf 513}, L127

\bibitem[\protect\astroncite{{Silverstone}}{2000}]{Silverstone}
{Silverstone}, M. 2000, {\em PhD Thesis\/}, UCLA

\bibitem[\protect\astroncite{{Song} {\em et~al.\/}}{2002}]{MF_TWA}
{Song}, I., {Bessell}, M.~S., \& {Zuckerman}, B. 2002, {\em \aap\/}, {\bf 385},
  862 (SBZ)

\bibitem[\protect\astroncite{{Song} {\em et~al.\/}}{2001}]{AVegas}
{Song}, I., {Caillault}, J.-P., {Barrado y Navascu{\' e}s}, D., \& {Stauffer},
  J.~R. 2001, {\em \apj\/}, {\bf 546}, 352 (S2001)

\bibitem[\protect\astroncite{{Song} {\em et~al.\/}}{2000}]{LateVegas}
{Song}, I., {Caillault}, J.-P., {Barrado y Navascu{\' e}s}, D., {Stauffer},
  J.~R., \& {Randich}, S. 2000, {\em \apjl\/}, {\bf 533}, L41 (S2000)

\bibitem[\protect\astroncite{{Song} {\em et~al.\/}}{2003}]{SZB03}
{Song}, I., {Zuckerman}, B., \& {Bessell}, M.~S. 2003, {\em \apj\/}, in press,
  Dec 10 (SZB)

\bibitem[\protect\astroncite{{Spangler} {\em et~al.\/}}{2001}]{Spangler}
{Spangler}, C., {Sargent}, A.~I., {Silverstone}, M.~D., {Becklin}, E.~E., \&
  {Zuckerman}, B. 2001, {\em \apj\/}, {\bf 555}, 932

\bibitem[\protect\astroncite{{Takeuchi} \& {Artymowicz}}{2001}]{Takeuchi}
{Takeuchi}, T. \& {Artymowicz}, P. 2001, {\em \apj\/}, {\bf 557}, 990

\bibitem[\protect\astroncite{{Wahhaj} {\em et~al.\/}}{2003}]{Wahhaj}
{Wahhaj}, Z., {Koerner}, D.~W., {Ressler}, M.~E., {Werner}, M.~W., {Backman},
  D.~E., \& {Sargent}, A.~I. 2003, {\em \apjl\/}, {\bf 584}, L27

\bibitem[\protect\astroncite{{Weinberger} {\em et~al.\/}}{2003}]{Weinberger03}
{Weinberger}, A.~J., {Becklin}, E.~E., \& {Zuckerman}, B. 2003, {\em \apjl\/},
  {\bf 584}, L33

\bibitem[\protect\astroncite{{Weinberger} {\em et~al.\/}}{2002}]{Weinberger02}
{Weinberger}, A.~J., {\em et~al.\/} 2002, {\em \apj\/}, {\bf 566}, 409

\bibitem[\protect\astroncite{{Wilner} {\em et~al.\/}}{2002}]{Wilner}
{Wilner}, D.~J., {Holman}, M.~J., {Kuchner}, M.~J., \& {Ho}, P.~T.~P. 2002,
  {\em \apjl\/}, {\bf 569}, L115

\bibitem[Wyatt(2003)]{Wyatt03} Wyatt, M.~C.\ 2003, ArXiv
Astrophysics e-prints, 8253

\bibitem[Wyatt \& Dent(2002)]{WD02} Wyatt, M.~C.~\& Dent,
W.~R.~F.\ 2002, \mnras, 334, 589

\bibitem[\protect\astroncite{{Zuckerman}}{2001}]{Zuckerman01}
{Zuckerman}, B. 2001, {\em \araa\/}, {\bf 39}, 549

\bibitem[\protect\astroncite{{Zuckerman} \& {Becklin}}{1993}]{Z93}
{Zuckerman}, B. \& {Becklin}, E.~E. 1993, {\em \apj\/}, {\bf 414}, 793

\bibitem[\protect\astroncite{{Zuckerman} {\em et~al.\/}}{1995}]{Z95}
{Zuckerman}, B., {Forveille}, T., \& {Kastner}, J.~H. 1995, {\em \nat\/}, {\bf
  373}, 494 (ZFK)

\bibitem[\protect\astroncite{{Zuckerman} \& {Song}}{2004}]{ARAA}
{Zuckerman}, B. \& {Song}, I. 2004, {\em \araa\/}, {\bf 42}, in press

\bibitem[\protect\astroncite{{Zuckerman} {\em
  et~al.\/}}{2001{\natexlab{a}}}]{bPic2}
{Zuckerman}, B., {Song}, I., {Bessell}, M.~S., \& {Webb}, R.~A.
  2001{\natexlab{a}}, {\em \apjl\/}, {\bf 562}, L87 (ZSBW)

\bibitem[\protect\astroncite{{Zuckerman} {\em
  et~al.\/}}{2001{\natexlab{b}}}]{ZSW}
{Zuckerman}, B., {Song}, I., \& {Webb}, R.~A. 2001{\natexlab{b}}, {\em \apj\/},
  {\bf 559}, 388 (ZSW)

\bibitem[\protect\astroncite{{Zuckerman} \& {Webb}}{2000}]{Z00}
{Zuckerman}, B. \& {Webb}, R.~A. 2000, {\em \apj\/}, {\bf 535}, 959 (ZW)
\end{thebibliography}
\end{document}